\documentclass[aps,pra,twocolumn,showpacs,superscriptaddress,preprintnumbers,floatfix,nofootinbib,]{revtex4-2}

\usepackage[pagebackref=true,colorlinks,citecolor=myred,linkcolor=myblue,urlcolor=myblue,linktocpage=true]{hyperref}

\usepackage{amsmath,amssymb}
\usepackage{epsfig}
\usepackage{hyperref}
\usepackage{color}
\usepackage{slashed}
\usepackage{placeins}
\usepackage{float}
\usepackage{amsfonts}
\usepackage{graphicx, rotating}
\usepackage{epstopdf}
\usepackage{latexsym}
\usepackage{rotating}
\usepackage{braket}
\usepackage{comment}
\usepackage{caption}
\usepackage{subcaption}
\usepackage[utf8]{inputenc}
\usepackage{hyperref}
\usepackage{siunitx}

\usepackage[export]{adjustbox}
\usepackage[font={small}]{caption}   

\usepackage[starfontserif]{starfont}

\usepackage{caption}
\usepackage{ragged2e}
\DeclareCaptionJustification{justified}{\justifying}
\counterwithin{equation}{section}

\DeclareSymbolFont{starfontsym}{OMS}{cmsy}{m}{n} 

\DeclareMathSymbol{\mathMoon}{\mathord}{starfontsym}{97}

\definecolor{myred}{rgb}{0.7, 0, 0}
\definecolor{myblue}{rgb}{0, 0, 0.7}
\definecolor{mygreen}{rgb}{0.04, 0.7, 0.5}
\definecolor{mygray}{rgb}{0.1, 0.1, 0.1}

\hypersetup{colorlinks,citecolor=myred,linkcolor=myblue,urlcolor=myblue,linktocpage=true}

\def\be   {\begin{equation}}   \def\ee   {\end{equation}}
\def\ba   {\begin{array}}      \def\ea   {\end{array}}
\def\bea  {\begin{eqnarray}}   \def\eea  {\end{eqnarray}}
\def\bean {\begin{eqnarray*}}  \def\eean {\end{eqnarray*}}

\def\bry{\begin{array}}
	\def\ery{\end{array}}

\newcommand{\skipnew}[1]{}

\numberwithin{equation}{section}

\def\mainz{Institute of Physics, Johannes Gutenberg-Universit\"at Mainz, 55099 Mainz, Germany}

\def\HIM{Helmholtz Institute Mainz, Staudingerweg 18, 55128
Mainz, Germany}

\def\Hebrew{Institute of Applied Physics, The Hebrew University of Jerusalem, Jerusalem 9190401, Israel}

\def\Racah{Racah Institute of Physics, The Hebrew University of Jerusalem, Jerusalem 9190401, Israel.}

\def\UOW{Department of Physics, University of Washington, Seattle, WA 98195-1560, USA}

\def\prinston{Princeton University, Department of Electrical and Computer Engineering, Princeton, New Jersey 08544, USA}

\def\UOWE{Electrical and Computer Engineering Department, University of Washington, Seattle, Washington 98105, USA}

\def\AIST{National Institute of Advanced Industrial Science and Technology (AIST), Tsukuba Central 2, 1-1-1, Umezono, Tsukuba, Ibaraki 305-8568, Japan}

\def\GSI{GSI Helmholtzzentrum für Schwerionenforschung GmbH, 64291 Darmstadt, Germany}

\def\ucb{Department of Physics, University of California, Berkeley, California 94720, USA}

\begin{document}

\date{\today}
\title{\Large\bfseries Magnetic Microscopy of Skyrmions in Magnetic Thin Films with Chiral Overlayers}


%

\author{Buddhika Hondamuni}
\email{hondamub@uni-mainz.de}
\affiliation{\mainz}
\affiliation{\HIM}

\author{Théo Balland}
\affiliation{\mainz}

\author{Fabian Kammerbauer}
\affiliation{\mainz}

\author{Ashish Moharana}
\affiliation{\mainz}

\author{Bindu}
\affiliation{\Hebrew}

\author{Amandeep Singh}
\affiliation{\Hebrew}

\author{Meital Ozeri}
\affiliation{\Hebrew}

\author{Shira Yochelis}
\affiliation{\Hebrew}

\author{Yossi Paltiel}
\affiliation{\Hebrew}

\author{Omkar Dhungel}
\affiliation{\mainz}
\affiliation{\HIM}

\author{Zeeshawn Kazi}
\affiliation{\prinston}

\author{Kai-Mei C. Fu}
\affiliation{\UOW}
\affiliation{\UOWE}

\author{Hideyuki Watanabe}
\affiliation{\AIST}

\author{Mathias Kl{\"a}ui}
\affiliation{\mainz}

\author{Arne Wickenbrock}
\affiliation{\mainz}
\affiliation{\HIM}
\affiliation{\GSI}

\author{Nir Bar-Gill}
\affiliation{\Hebrew}
\affiliation{\Racah}

\author{Angela Wittmann}
\affiliation{\mainz}

\author{Dmitry Budker}
\email{budker@uni-mainz.de}
\affiliation{\mainz}
\affiliation{\HIM}
\affiliation{\GSI}
\affiliation{\ucb}

\begin{abstract}
Topologically nontrivial magnetic textures such as skyrmions offer promising opportunities for spintronic applications. In recent years, it has been shown that the magnetic properties of layered materials can be affected by depositing chiral molecules on the surface, while the influence of chiral overlayers on skyrmion properties such as their stability and interactions remains largely unexplored. To address this challenge, we employ wide-field nitrogen-vacancy (N$V$) magnetometry to directly image skyrmions in chiral-molecule-functionalized magnetic thin films, enabling quantitative mapping of magnetic stray fields over extended areas under ambient conditions. Using pixel-resolved optically detected magnetic resonance (ODMR) combined with controlled magnetic fields, we reproducibly nucleate and probe skyrmion states in CoFeB ferromagnetic samples, enabling quantitative investigation of their properties. We find evidence for enantioselective and magnetic-field-polarity-dependent modifications of skyrmion diameter, spacing, and shape, pointing to a possibility of molecular control of topological spin textures via magneto-chiral coupling.
\end{abstract}

\maketitle

\section{Introduction}

Magnetic skyrmions are nanoscale topological spin textures~\cite{fert2013skyrmions} that have attracted considerable interest for next-generation spintronic technologies due to their stability~\cite{bogdanov1994thermodynamically}, small size~\cite{heinze2011spontaneous}, and efficient manipulation by external stimuli such as magnetic fields and electric currents~\cite{skyrme1961non,zhang2023magnetic,everschor2018perspective}. Understanding how to control these topologically nontrivial spin configurations is therefore of both fundamental and technological importance.

Thin film magnetic systems support nanoscale domain structures and skyrmions that are particularly relevant for high-density magnetic storage applications~\cite{nagaosa2013topological,fert2017magnetic,reichhardt2022statics}. Their magnetic properties can be engineered through interface design, multilayer stacking~\cite{jaiswal2019tuning}, and strain~\cite{shen2022strain}, enabling control over domain formation, stability, and dynamics~\cite{dohi2022thin}. Consequently, magnetic thin films provide a versatile platform for investigating controllable topological spin textures.

An emerging direction for controlling magnetic textures involves the interaction between magnetic materials and chiral molecules. The chirality-induced spin selectivity (CISS) effect demonstrates that electron spin polarization can depend on molecular handedness~\cite{naaman2019chiral}. Chiral molecules have been shown to influence magnetic behavior~\cite{ben2017magnetization}, chemical reactivity, and spin-dependent electron transport. In particular, enantioselective adsorption of chiral molecules on ferromagnetic surfaces has been shown to depend on the substrate magnetization direction, indicating a coupling between molecular chirality and magnetic order~\cite{safari2024enantioselective}. Such interactions suggest that molecular chirality may modify magnetic domain configurations and stabilize specific spin textures, offering a route toward functional hybrid molecular--magnetic systems~\cite{kapon2024effects}.

Recent experiments using magneto-optic Kerr effect (MOKE) microscopy explore this interaction in magnetic multilayer functionalized with chiral polypeptides. These studies demonstrate that adsorption of chiral molecules can modify magnetic properties such as coercivity, domain pinning, and magnetic anisotropy, and can influence the formation and stability of magnetic domains~\cite{kapon2023MOKE,kapon2024effects}. However, MOKE imaging primarily probes magnetization contrast and does not provide direct, quantitative information about the stray magnetic fields generated by the underlying spin textures, limiting access to the local magnetic energy landscape.

Understanding how chiral molecules influence skyrmion properties therefore requires experimental techniques capable of quantitatively mapping stray magnetic fields associated with magnetic textures over extended areas. Scanning single N$V$ magnetometry offers nanoscale spatial resolution and high magnetic sensitivity, but the sequential scanning approach typically restricts the field of view (FOV) to a few micrometers and leads to long acquisition times when imaging extended magnetic textures.

Wide-field N$V$ magnetometry provides a complementary approach by enabling parallel optical detection of ODMR from ensembles of near-surface N$V^{-}$ centers in diamond. This technique allows quantitative imaging of static stray magnetic fields with sub-micrometer spatial resolution over fields of view spanning tens of micrometers under ambient conditions~\cite{rondin2014magnetometry,levine2019principles}. Previous studies combining wide-field N$V$ sensing with magneto-optic measurements have visualized stripe domains in multilayer ferromagnet~\cite{lenz2021imaging}, while complementary experiments using N$V$ magnetometry and related nanoscale probes have revealed detailed properties of skyrmions and domain structures in ultra-thin magnetic films and multilayers~\cite{rana2020room,velez2022current,gross2018skyrmion,dovzhenko2018magnetostatic,jenkins2019single}. More broadly, N$V$ magnetometry has been applied to visualize stray magnetic fields in ferromagnetic films, magnetic nanoparticles, biological systems, and superconducting vortices, as well as to reconstruct current densities in electronic and mesoscopic devices~\cite{simpson2016magneto,le2013optical,schlussel2018wide,nowodzinski2015nitrogen,lillie2019imaging,casola2018probing,morgenbesser2021cation,shaji2024microwave,sengottuvel2025microwave}.

In this work, we use wide-field continuous-wave (CW) N$V^{-}$ center magnetometry to quantitatively image the stray magnetic fields of skyrmions in multilayer ferromagnet functionalized with chiral $\alpha$-helix poly-alanine molecules under ambient conditions. By directly comparing molecule-functionalized and non-functionalized regions of the same sample, we access spatial variations in skyrmion size, spacing, and morphology through their stray-field signatures, enabling a quantitative investigation of how molecular chirality modifies the local magnetic energy landscape. This approach provides information that is not accessible with conventional MOKE imaging and complements scanning probe techniques by combining quantitative field reconstruction with a large FOV. In the present implementation, the magnetic image is acquired in approximately 10 min (corresponding to about 0.4\,s per frequency point). While the acquisition time could be reduced by decreasing the number of sampled frequency points, we intentionally employ dense spectral sampling with averaging to enhance the signal-to-noise ratio and enable robust pixel-wise ODMR fitting, thereby improving the reliability and quality of the reconstructed magnetic field maps.

\section{Topological defects in ordered magnets}
\label{sec:Skyrmions}

Magnetic materials undergo a transition from a paramagnetic state with randomly oriented spins to an ordered phase with finite magnetization below a critical temperature. This spontaneous symmetry breaking establishes long-range magnetic order and enables the formation of spatially structured spin textures, including topological defects~\cite{lancaster2019skyrmions}. Owing to their nontrivial topology and nanoscale size, skyrmions are of particular interest for both fundamental studies and spintronic applications. Skyrmion is characterized by a continuous rotation of the magnetization that wraps the unit sphere, leading to a quantized topological charge (winding number)~\cite{buttner2016magnetic,everschor2018perspective,dohi2022thin}. Their stability results from the competition among exchange interaction, magnetic anisotropy, Zeeman energy, and the Dzyaloshinskii--Moriya interaction (DMI), which favors chiral spin textures. Depending on whether the DMI is interfacial or bulk in origin, Néel-type or Bloch-type skyrmions can form. These configurations may appear either as metastable states or as global energy minima, depending on material parameters, applied field, and geometry~\cite{roessler2006spontaneous,muhlbauer2009skyrmion,yu2010real,seki2012observation,nayak2017magnetic,heinze2011spontaneous,romming2013writing}.

\section{Magneto-Chiral Effect}
\label{sec:CISS}

Control of skyrmions through external magnetic fields, currents and interfacial engineering are well established~\cite{everschor2018perspective}. An emerging approach exploits molecular chirality, which represents \emph{true} chirality in the symmetry sense and introduces structural handedness. In contrast, magnetization corresponds to a form of \emph{false} chirality, and their coexistence enables hybrid magneto--chiral systems~\cite{kapon2024effects,barron1986true}. Chiral molecules, which are not super-imposable on their mirror images, can influence spin-dependent processes through the CISS effect, where electron transmission becomes spin selective~\cite{naaman2015spintronics,naaman2018chirality,bloom2024chiral}. While CISS is commonly studied in transport geometries, chiral molecular overlayers on magnetic thin films can induce enantiomer-dependent modifications of magnetic properties~\cite{moharana2025chiral}. Within a symmetry-based framework, the broken spatial inversion symmetry introduced by molecular chirality can modify effective magnetic coupling to the magnetic layer, and thus affect the spin textures. Consistent with this picture, enantioselective adsorption of chiral molecules has been shown to modify magnetic properties at the surfaces of ferromagnetic materials, including coercivity, magnetic anisotropy, and domain behavior~\cite{kapon2023MOKE}. Here we interpret the handedness and polarity-dependent changes in skyrmion size, spacing, and shape as a manifestation of such magneto-chiral coupling.

\begin{figure*}
    \centering
    \includegraphics[width=1\textwidth]{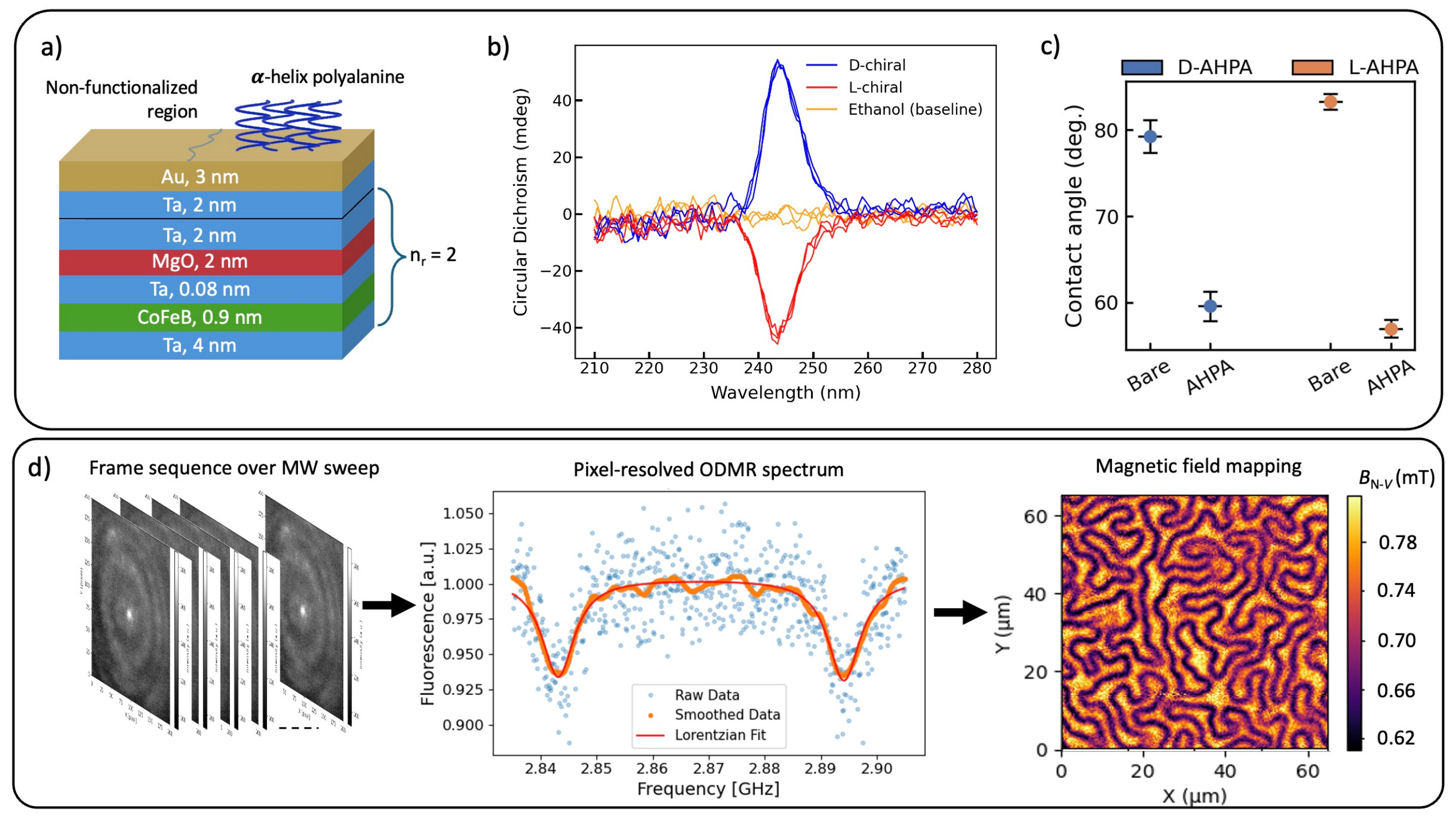}
    \caption{\textbf{(a) Sample design:} Schematic of the multilayer magnetic stack with a $\sim$3\,nm Au capping layer, where chiral molecules are deposited on only half of the surface, leaving an adjacent non-functionalized region for direct comparison. \textbf{Molecular characterization:} \textbf{(b)} Circular dichroism (CD) spectra of D- and L-AHPA solutions measured in the UV range (210–280\,nm), showing mirror-symmetric extrema near $\sim$240–245\,nm that confirm opposite molecular handedness; the ethanol reference remains close to zero. \textbf{(c)} Contact angle measurements for bare and molecule-functionalized regions, demonstrating the change in surface wettability after adsorption of D-AHPA and L-AHPA molecules. \textbf{(d) N$V$ magnetometry workflow:} A sequence of photoluminescence frames acquired during a microwave frequency sweep yields pixel-resolved ODMR spectra for the FOV, each spectrum is fitted with a lorentzian function to determine the resonance positions and splitting, enabling quantitative reconstruction of the stray magnetic field map.}
    \label{fig:cd_spectra}
\end{figure*}

\section{Magnetic samples and chiral molecules}

The materials studied here are thin films with perpendicular magnetic anisotropy (PMA), in which the magnetization preferentially aligns normal to the film plane. They form complex domain patterns under applied bias fields. The multilayer films were deposited by DC magnetron sputtering using a Singulus Rotaris deposition tool with the structure
$\mathrm{Si}/\allowbreak
\mathrm{Si}(100)/\allowbreak
\mathrm{Ta}(4)/\allowbreak
[\mathrm{Co}_{20}\mathrm{Fe}_{60}\mathrm{B}_{20}(0.9)/\allowbreak
\mathrm{Ta}(0.08)/\allowbreak
\mathrm{MgO}(2)/\allowbreak
\mathrm{Ta}(2)]_{2}/\allowbreak
\mathrm{Ta}(2)$,
where numbers in parentheses denote the layer thickness in nanometers [Fig.\,\ref{fig:cd_spectra}\,(a)]. This stack supports PMA and stabilizes magnetic domains suitable for skyrmion formation. As the chiral overlayer, we employ $\alpha$-helix poly-alanine (AHPA) polypeptide, which is known to induce strong domain pinning and a localized increase in coercive field~\cite{kapon2024effects}. The molecules have a purity of approximately 95\%, terminate with a surface –OH group, and possess a calculated molecular mass of 3008.6\,g/mol. The molecular solution was prepared at a concentration of 1\,mM in ethanol.

\begin{figure*}
    \centering
    \includegraphics[width=0.91\textwidth]{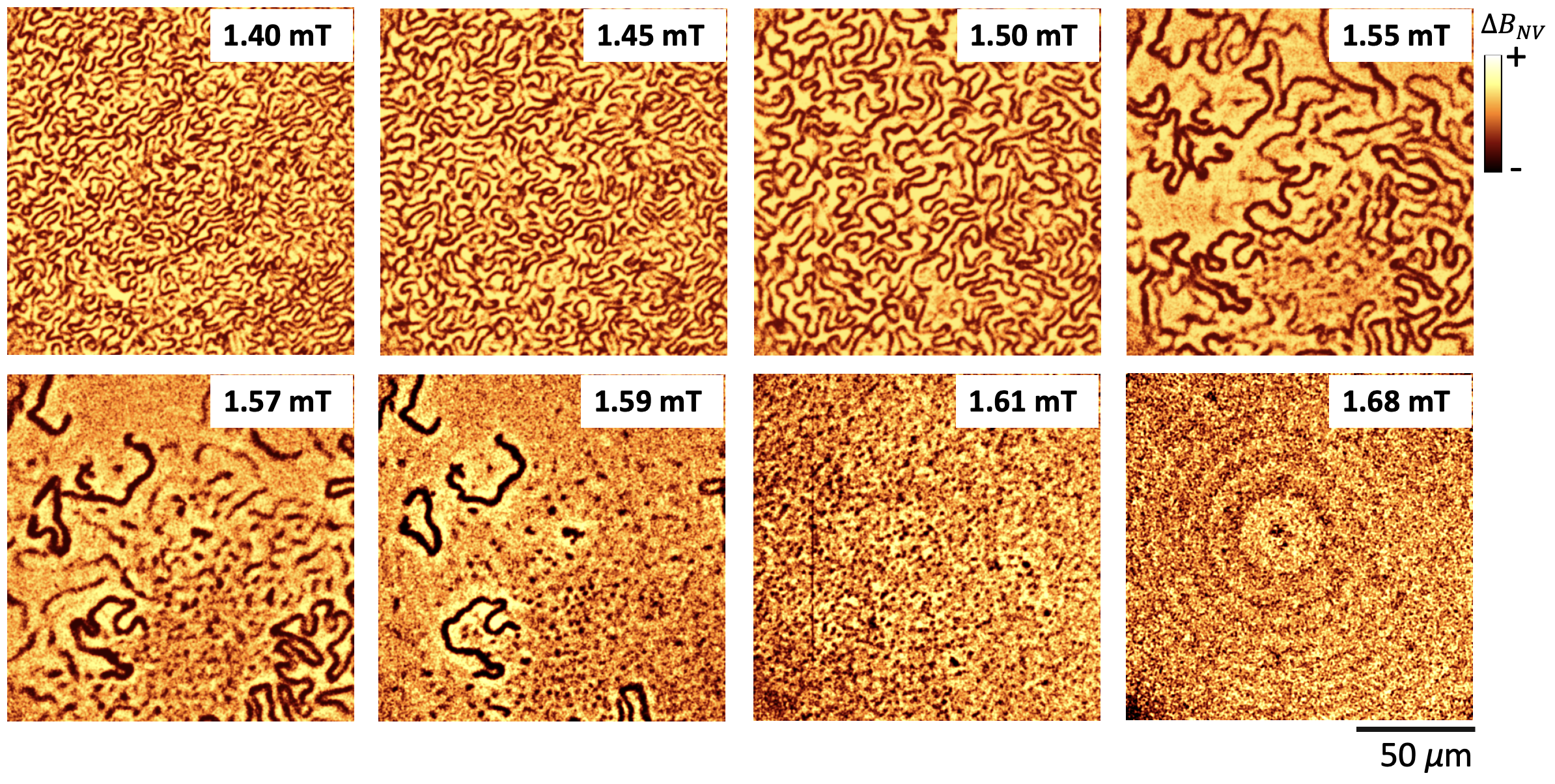}
    \caption{\textbf{Evolution of magnetic domains to skyrmion phase in CoFeB thin film under increasing $B_{z}$ field.} Background-subtracted N$V$ magnetic field maps (\(\Delta B_{\mathrm{NV}}\)) reveal the progressive deformation and coalescence of skyrmions together with stripe domains by increasing OOP field. The magnetic contrast is normalized by the standard deviation $\sigma$ ($\sim 1.15\,\mu\mathrm{T}$) of the field distribution, so that the color scale reflects statistically significant deviations from the background. The faint concentric ring pattern originates from optical interference caused by reflections within the diamond.}
    \label{fig:skyrmion_evolution}
\end{figure*}

\section{Experimental Methods}

To isolate the influence of molecular chirality while minimizing sample-to-sample variations, we use fabricated samples containing adjacent non-functionalized and molecule-functionalized regions. Each sample was first coated with a protective poly-methyl methacrylate (PMMA) layer. One half of the surface was then masked with Kapton tape, and the exposed region was cleaned by dissolving the PMMA in acetone, producing a boundary between protected and unprotected areas. After removing the mask, the samples were immersed overnight in solutions containing the chiral molecules, enabling selective adsorption on the exposed region. The molecules bind to the gold surface via chemisorption of the terminal cysteine thiol group (Au–S bond) and adopt a tilted orientation with an angle of approximately $46^\circ$ relative to the surface normal. The resulting self-assembled monolayer provides dense surface coverage with an intermolecular spacing of approximately 12 Angstrom and remains stable against brief acetone exposure at room temperature~\cite{nguyen2022cooperative}. Finally, the remaining PMMA was removed by a short acetone bath ($\sim$2\,min), yielding samples with adjacent regions with and without chiral molecular overlayers.

The chirality of the molecular solutions was confirmed by circular dichroism spectroscopy [Fig.\,\ref{fig:cd_spectra}\,(b)], which shows opposite sign responses for the D and L enantiomers, whereas the ethanol baseline remains featureless, confirming the absence of intrinsic chirality. Contact angle measurements confirm successful molecular adsorption [Fig.\,\ref{fig:cd_spectra}\,(c)]. For the D-AHPA sample, the average contact angle decreases from $79.25^\circ$ (non-functionalized) to $59.56^\circ$ after functionalization, while for the L-AHPA sample it decreases from $83.33^\circ$ to $56.97^\circ$. The reduced contact angles indicate increased surface wettability consistent with the presence of the molecular overlayer. With the molecular handedness thus established, the effect of chiral adsorption on skyrmion properties was investigated by directly comparing the functionalized and non-functionalized regions on the same sample under identical experimental conditions.

Skyrmions are nucleated by applying a combination of out-of-plane (OOP) and in-plane (IP) magnetic fields, where the IP component breaks symmetry and enables nucleation in systems with insufficient intrinsic DMI. The resulting skyrmions are expected to be Néel-type, stabilized by interfacial DMI in the multilayer structure, a widely used platform for hosting relatively large skyrmions in PMA films~\cite{fert2013skyrmions,nagaosa2013topological}. 

Spatially resolved ODMR imaging is used to reconstruct the stray magnetic field produced by PMA magnetic thin films. The N$V$ centers are optically polarized into the $m_s=0$ state using CW 532\,nm illumination (approximately 70\,mW), while a microwave (MW) field drives spin transitions. The spin-state-dependent photoluminescence (PL) is collected by the objective and imaged onto the sCMOS camera. During acquisition, the MW frequency is swept over a bandwidth centered at the N$V$ zero-field splitting at $2.87\,\mathrm{GHz}$ to record the full ODMR resonance [Fig.\,\ref{fig:cd_spectra}\,(d)]. For each MW frequency step, a fluorescence image is acquired synchronously, producing a stack of $N$ number of images corresponding to discrete points of the ODMR spectrum. Each image is normalized to a reference frame acquired at a fixed off-resonance frequency, to suppress common-mode variations during the measurement~\cite{scholten2021widefield}. The frequency sweep is repeated multiple times to improve the signal-to-noise ratio, with an effective integration time of 100\,ms per frequency point. An ODMR spectrum is extracted for each pixel, and the resonance positions are determined by pixel-wise Lorentzian fitting. The spatial distribution of resonance shifts is subsequently converted into a two-dimensional magnetic field map, yielding the stray-field component $B_{\mathrm{NV}}$ over the FOV.

\begin{figure}[b]
\centering
\includegraphics[scale=0.29]{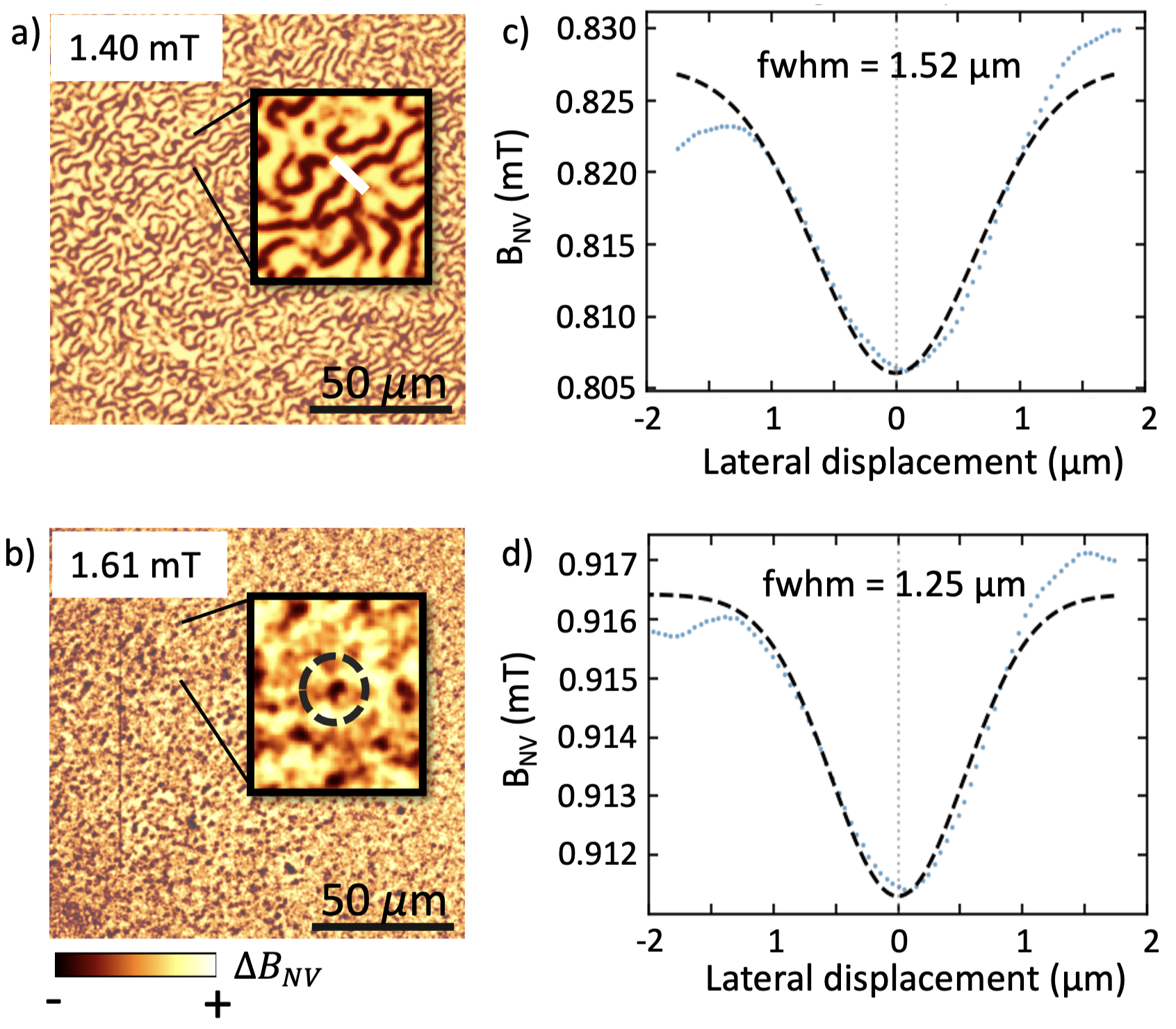}
\caption{\textbf{Magnetic field profiles of DWs and skyrmions.}
\textbf{(a, b)} Insets display magnified regions; the white line in (a) indicates the path used to extract the line profile across a DW, while the dashed circle in (b) marks a representative skyrmion used for analysis. \textbf{(c, d)} Corresponding stray magnetic field $B_{\mathrm{NV}}$ as a function of lateral displacement across a DW (c) and a skyrmion (d), centered at 0. Dashed lines represent Gaussian fits used to extract the full width at half maximum (FWHM).}
\label{fig:FWHM}
\end{figure}

\section{Results and discussion}

Figure\,\ref{fig:skyrmion_evolution} presents N$V$ stray-magnetic field maps of magnetic domain configurations, illustrating the evolution from stripe domains to isolated skyrmions under an applied OOP magnetic field. The measured signal corresponds to the projection of the magnetic field along the N$V$ axis, so that the reconstructed maps represent a weighted combination of IP and OOP stray-field components. For thin films with PMA, the stray field above the sample is dominated by the OOP component, such that $B_{\mathrm{NV}}$ provides a reliable proxy for $B_z$ up to a geometric scaling factor. We note that no full vector-field or magnetization reconstruction is performed in this work. Such reconstruction would require measurements along multiple N$V$ orientations or inverse modeling, which is beyond the scope of the present study. Instead, our analysis focuses on comparative trends between molecule-functionalized and reference regions measured under identical conditions.

\begin{figure*}
    \centering
    \includegraphics[width=0.9\textwidth]{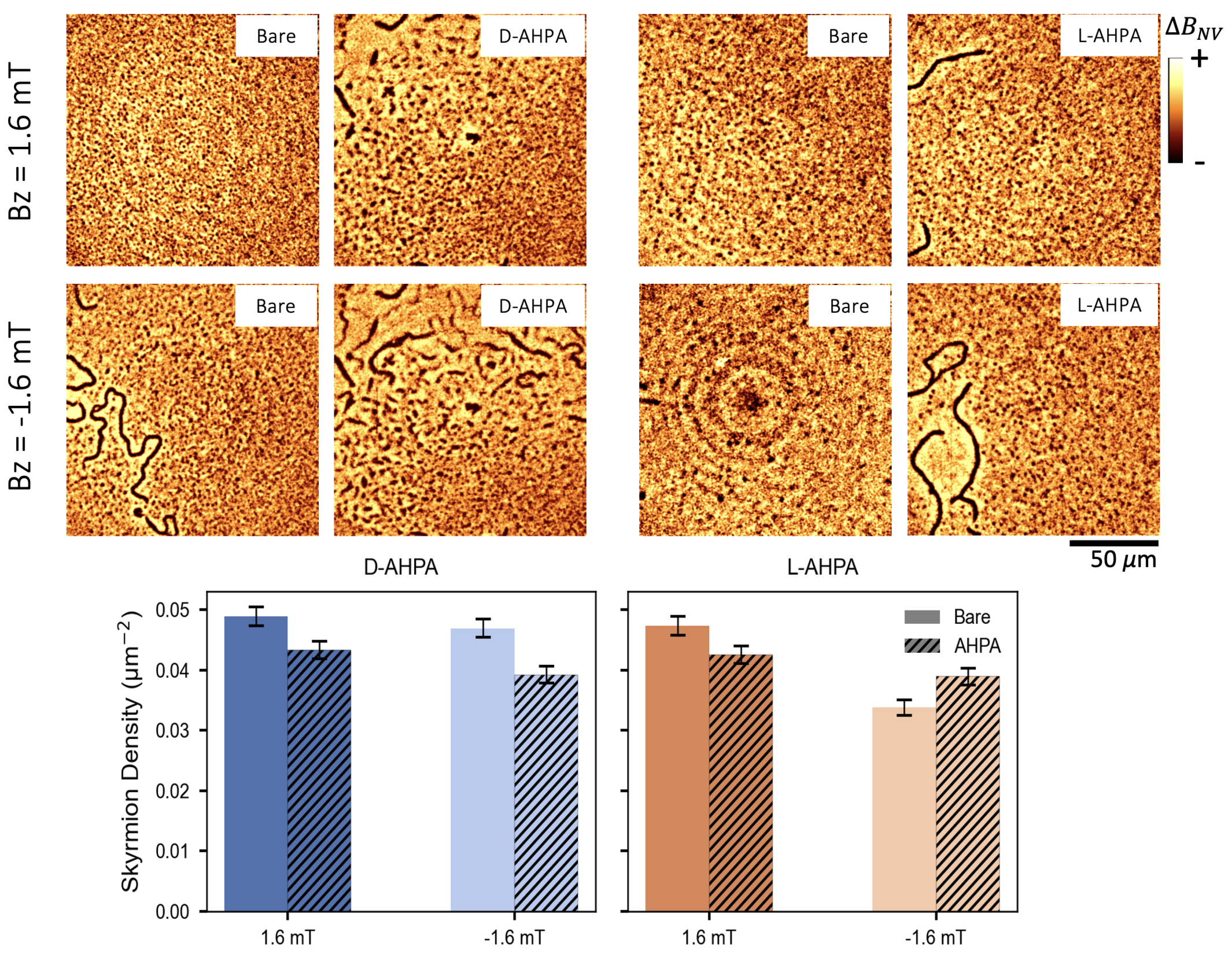}
    \caption{\textbf{Chiral-molecule-induced modification of skyrmion textures.} Stray magnetic field images of skyrmions in bare, D-AHPA, and L-AHPA functionalized regions measured under $B_z = \pm 1.6\,\mathrm{mT}$ following the application of a $\sim 30\,\mathrm{mT}$ IP magnetic pulse. While isolated skyrmions dominate the observed textures, a small fraction of residual labyrinthine domains persists. The lower panels show the corresponding skyrmion density within the FOV.}
    \label{fig:chirality_comparison}
\end{figure*}

\begin{figure*}
    \centering
    \includegraphics[width=0.8\textwidth]{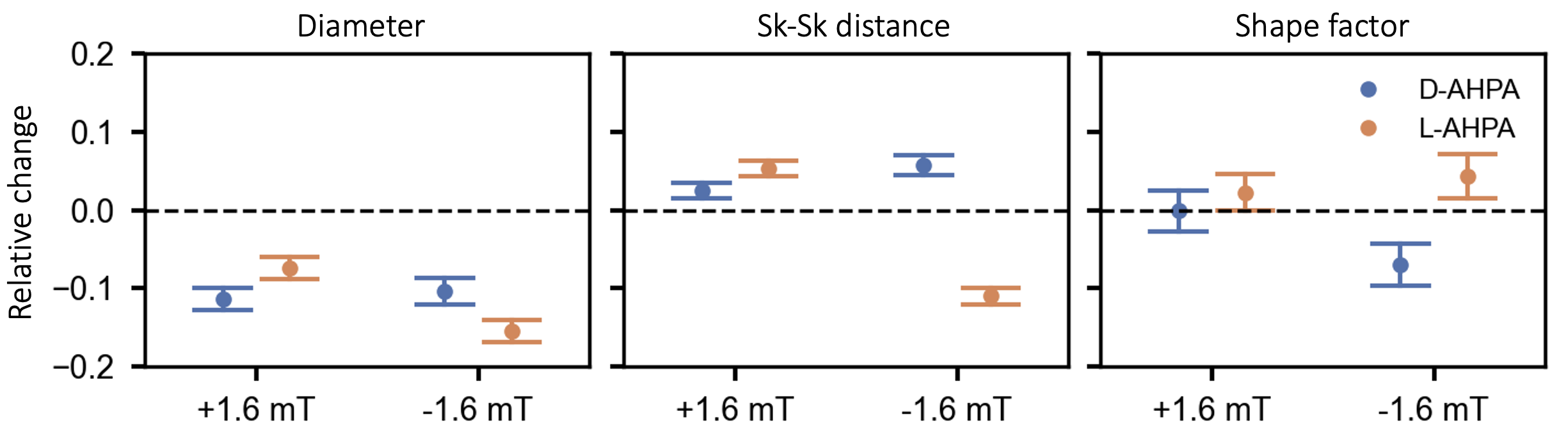} 
    \caption{\textbf{Statistical distributions of relative changes:} in skyrmion diameter, $d_{Sk-Sk}$, and shape factor for D- and L-AHPA functionalized regions relative to the non-functionalized regions under $B_z=\pm1.6\,\mathrm{mT}$.}
    \label{fig:stat_comparison}
\end{figure*}

Magnetic stripes are nucleated by applying a brief ($\sim$1\,s) IP magnetic field pulse of approximately 30\,mT using a laterally movable permanent magnet while maintaining a constant OOP bias field. This pulse destabilizes the stripe domain walls (DWs), enabling the formation of skyrmions. After removal of the IP field, the resulting magnetic configuration depends on the OOP field, intrinsic sample parameters, and temperature. The combined use of controlled OOP bias and transient IP excitation provides a reliable method for generating the skyrmion phase.

Following this preparation, the domain morphology undergoes a systematic evolution as the OOP field increases. At low fields ($B_z \approx 1.40$--$1.50\,\mathrm{mT}$), the film exhibits a dense labyrinthine stripe-domain phase. At $B_z \approx 1.55\,\mathrm{mT}$, the stripes narrow and fragment, producing a mixed state of broken stripes and emerging circular textures that signal the onset of skyrmion nucleation. At higher fields ($B_z \approx 1.57$--$1.59\,\mathrm{mT}$), isolated skyrmions dominate the magnetic contrast, appearing as individual circular textures embedded in a nearly uniform background and stabilized at defect-induced pinning sites commonly present in thin magnetic films~\cite{reichhardt2022statics}. When the field reaches $B_z \approx 1.61\,\mathrm{mT}$, most stripe domains vanish and only isolated skyrmions remain, indicating a skyrmion-dominated phase close to magnetic saturation. At $B_z \approx 1.68\,\mathrm{mT}$, the magnetic contrast diminishes as the film approaches magnetic saturation, and the fluorescence signal becomes nearly uniform. Such field-driven transitions from stripe domains to isolated skyrmions and ultimately to the saturated state are expected in PMA thin films and arise from the competition between exchange, magnetic anisotropy, dipolar interactions, and Zeeman energy~\cite{nagaosa2013topological,fert2013skyrmions}.

To quantify the characteristic length scales associated with these field-driven textures, we extract line profiles of the stray magnetic field $B_{\mathrm{NV}}$ across representative features [Fig.\,\ref{fig:FWHM}]. The resulting profiles provide a direct measure of the lateral extent of the magnetic textures through their FWHM. In the stripe-domain regime, the broader field minima reflect the extended nature of the domains, whereas in the skyrmion regime the profiles become more localized, indicating reduced characteristic size. Although the measured $B_{\mathrm{NV}}$ represents a convolution of the underlying magnetization with the dipolar stray-field response \cite{tetienne2015nature}, the extracted FWHM captures the relative size evolution of the magnetic textures. This quantitative approach enables direct comparison of DW and skyrmion dimensions.

To evaluate the influence of molecular chirality on skyrmion properties, we performed a statistical analysis of the skyrmion diameter, nearest-neighbor skyrmion--skyrmion (Sk--Sk) distance, and shape factor for non-functionalized and molecule-functionalized regions under opposite magnetic field polarities ($B_z = \pm 1.6\,\mathrm{mT}$). Because the magnetic field is reconstructed from the splitting of the two ODMR resonances, the measurement yields the magnitude of the N$V$-projected magnetic field; therefore the sign of the stray field is not resolved and skyrmions appear with similar contrast for both field polarities.

The resulting statistical distributions are summarized in Fig.\,\ref{fig:chirality_comparison}. For D-AHPA, the skyrmion density in the non-functionalized (bare) region is higher than in the molecule-functionalized region at both magnetic field polarities. A slight overall reduction in density is observed upon reversing the field from $B_z = +1.6\,\mathrm{mT}$ to $B_z = -1.6\,\mathrm{mT}$, while the relative difference between bare and functionalized regions remains consistent. For L-AHPA, the skyrmion density at $B_z = +1.6\,\mathrm{mT}$ is lower in the molecule-functionalized region compared to the bare region. In contrast, at $B_z = -1.6\,\mathrm{mT}$, the skyrmion density becomes higher in the functionalized region, indicating a reversal in behavior and an opposite trend compared to D-AHPA. Overall, these results reveal an enantioselective and polarity-dependent modification of skyrmion stability, where the two enantiomers produce qualitatively different responses under magnetic field reversal. It is important to note that near zero magnetic field, skyrmions tend to collapse into stripe-like domains. Consequently, when reversing the magnetic field polarity, the nucleation procedure must be repeated. Small variations in the IP nucleation conditions may therefore lead to differences in the measured skyrmion density between polarities and sample regions.

Figure\,\ref{fig:stat_comparison} presents the statistical distributions of relative changes in skyrmion properties for each enantiomer under opposite magnetic field polarities, referenced to the corresponding non-functionalized regions, quantified using the mean and standard error of the mean (SEM). Skyrmions are identified from the stray-field images using a segmentation algorithm based on a U-Net architecture, where masks are generated by optimizing intensity thresholds and Gaussian filtering parameters to accurately capture skyrmion cores and boundaries. The resulting masks are used to extract skyrmion positions and geometrical properties for further analysis. 

For D-AHPA, the skyrmion diameter distributions at $B_z = \pm 1.6\,\mathrm{mT}$ are largely comparable, indicating only a weak dependence on magnetic field polarity. In contrast, for L-AHPA, the diameter distribution at $+1.6\,\mathrm{mT}$ is shifted toward larger values compared to $-1.6\,\mathrm{mT}$, revealing a more pronounced polarity-dependent behavior than observed for D-AHPA. 

The skyrmion--skyrmion distance ($d_{Sk-Sk}$), defined as the nearest-neighbor center-to-center separation obtained from Delaunay triangulation of segmented skyrmion centroids~\cite{de2008computational}, reflects the distribution of pinning sites in a disordered landscape. For D-AHPA, the skyrmion spacing distribution at $B_z = +1.6\,\mathrm{mT}$ is shifted toward lower values compared to $B_z = -1.6\,\mathrm{mT}$. This indicates reduced nearest-neighbor distances under positive field, corresponding to a locally denser skyrmion arrangement. For L-AHPA, the distribution at $+1.6\,\mathrm{mT}$ is centered close to zero, while at $-1.6\,\mathrm{mT}$ it shifts toward negative values, indicating a reduction in nearest-neighbor distances under negative field. The magnitude of this shift is slightly larger than in the D-AHPA case, suggesting a stronger compression of skyrmion spacing. 

The skyrmion shape factor, defined as ${\varnothing_H - \varnothing_A}/{\varnothing_H}$, where ${\varnothing_H}$ is the diameter of the smallest convex hull
enclosing the skyrmion and ${\varnothing_A}$ is the mean diameter, quantifies deviations from circularity, with larger values indicating increased distortion. For D-AHPA, the distribution at $B_z = +1.6\,\mathrm{mT}$ is shifted toward higher values compared to $-1.6\,\mathrm{mT}$, indicating more distorted skyrmions under positive field. In contrast, L-AHPA exhibits a weaker polarity dependence. 

\begin{figure}[h]
\centering
\includegraphics[scale=0.33]{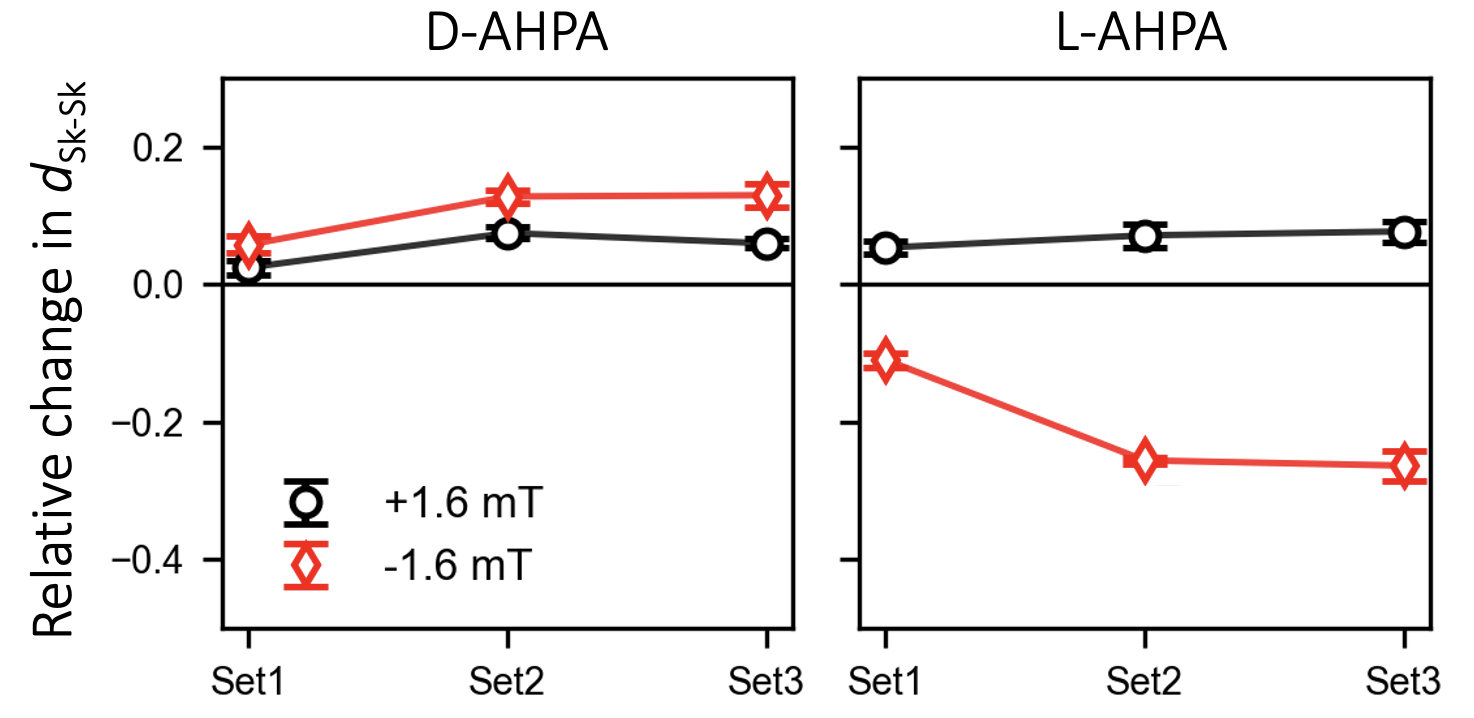}
\caption{\textbf{Relative change in $d_{Sk-Sk}$:} for D-AHPA and L-AHPA across three independent datasets. Solid lines connect the mean values for each dataset. Error bars represent the SEM.}
\label{fig:median_IQR}
\end{figure}

Figure\,\ref{fig:median_IQR} summarizes the statistical distributions of the relative change in $d_{Sk-Sk}$ across three independent data sets. Among the three quantities, the $d_{Sk-Sk}$ exhibits the most consistent and systematic variation with magnetic field polarity for both enantiomers. For D-AHPA, the mean $d_{Sk-Sk}$ is generally larger under $-1.6\,\mathrm{mT}$ compared to $+1.6\,\mathrm{mT}$, whereas for L-AHPA the opposite trend is observed, with reduced spacing under negative field. This polarity-dependent and enantioselective behavior indicates a modification of the effective skyrmion--skyrmion interaction and pinning landscape, leading to a compression or expansion of the skyrmion ensemble depending on molecular chirality and field direction.

The observed variations in skyrmion diameter, spacing, and shape can be understood in terms of modifications to the underlying magnetic energy balance. In thin films, these properties are governed by the interplay between interfacial DMI, PMA, dipolar interactions, and defect-induced pinning~\cite{rohart2013skyrmion,bogdanov1994thermodynamically,heinze2011spontaneous}. Changes in DMI and anisotropy directly influence the domain-wall energy, thereby affecting skyrmion size, while dipolar repulsion and pinning determine their spatial arrangement and local ordering. In addition, disorder and interfacial inhomogeneities can distort skyrmion shapes and modify their stability. The polarity-dependent and enantioselective trends observed here therefore indicate that chiral molecular adsorption alters these competing energy contributions at the interface, leading to measurable changes in skyrmion morphology and distribution.

\section*{Conclusions}

We demonstrate wide-field N$V$ magnetometry of chiral-molecule-functionalized magnetic thin films with perpendicular magnetic anisotropy, enabling direct microscale imaging of skyrmion stray-field textures under identical experimental conditions. By combining pixel-resolved optically detected magnetic resonance imaging with controlled out-of-plane and in-plane magnetic fields, skyrmion states are reproducibly nucleated and quantitatively analyzed.

The statistical analysis reveals systematic, enantioselective, and polarity-dependent modifications of the skyrmion configuration, indicating that molecular chirality alters the interfacial magnetic energy landscape. The observed asymmetry under magnetic field reversal suggests modifications of key interfacial parameters, including magnetic anisotropy, Dzyaloshinskii--Moriya interaction, and defect-induced pinning. In our system, molecular adsorption on a single surface introduces structural inversion asymmetry, providing a mechanism through which chirality couples to the interfacial spin structure. The reduced enantioselective contrast observed in thicker multilayer stacks further indicates that bulk dipolar interactions and disorder-driven pinning can partially suppress interfacial magneto-chiral effects.

Among the extracted observables, the nearest-neighbor skyrmion spacing emerges as the most robust indicator of chirality- and polarity-dependent behavior, while variations in other parameters remain more susceptible to local disorder and pinning. This highlights the importance of collective metrics for capturing systematic changes in skyrmion configurations.

These findings establish molecular chirality as an effective tuning parameter for controlling topological spin textures in hybrid molecular--magnetic systems. More broadly, they demonstrate wide-field N$V$ magnetometry as a powerful platform for investigating magneto-chiral interactions. In the present work, our primary objective is to demonstrate the capability of this technique to resolve differences between molecule-functionalized and reference regions. Extending this approach to larger statistical ensembles and multiple regions of the sample will be an important direction for future studies.
\label{sec:conclusion}

\begin{figure*}
    \centering
    \includegraphics[width=0.8\textwidth]{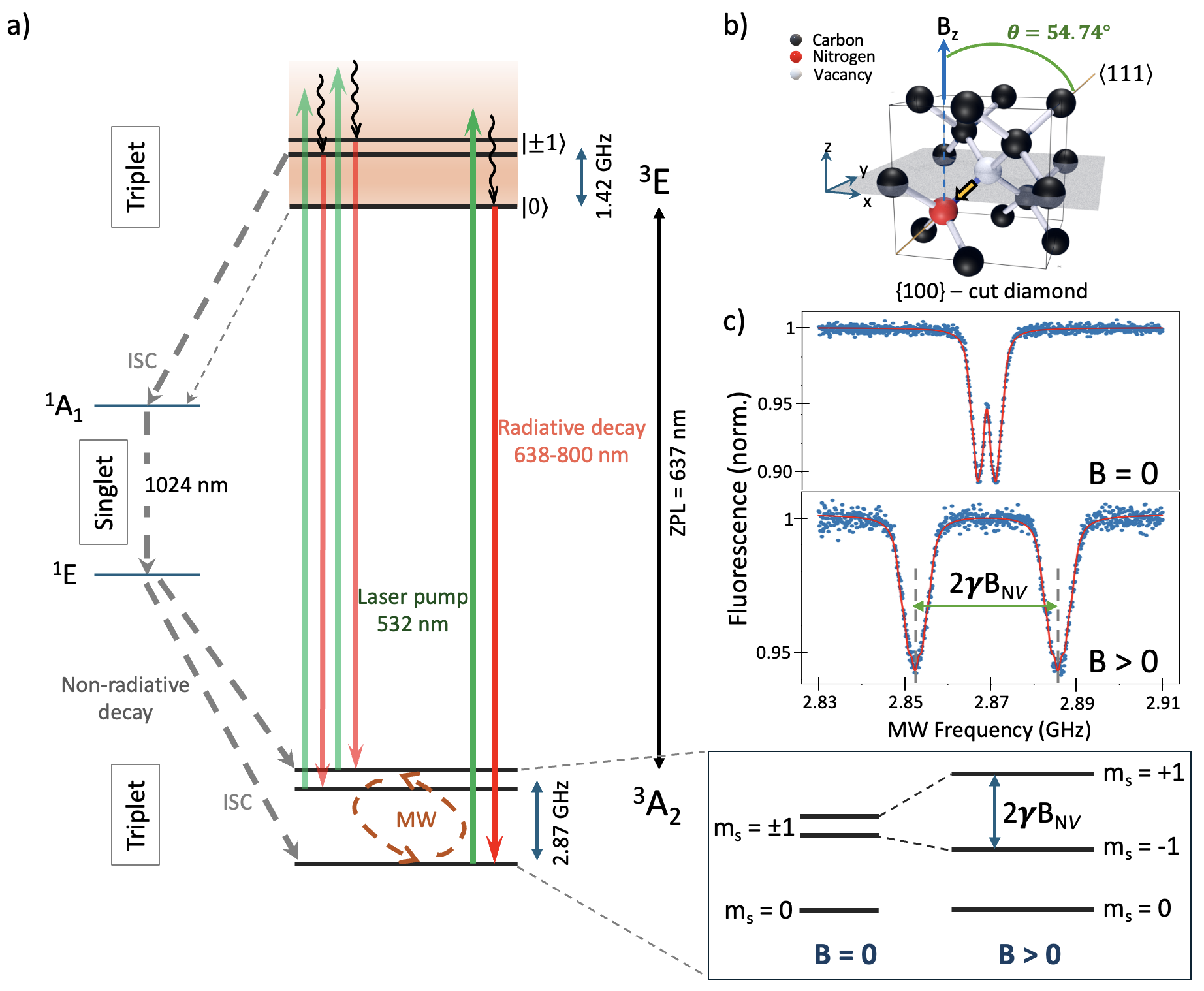}
\caption{\textbf{(a) Energy-level diagram of the N$V^{-}$ center.} The N$V$$^{-}$ center has a spin-1 ground-state triplet, $^{3}\mathrm{A}_{2}$; within this manifold the $m_{S}=0$ and $m_{S}=\pm1$ sublevels are split by the zero-field splitting $D \approx 2.87\,\mathrm{GHz}$ at room temperature (this splitting is internal to the ground triplet). The center also has an excited triplet state, $^{3}\mathrm{E}$; optical transitions between $^{3}\mathrm{A}_{2}$ and $^{3}\mathrm{E}$ (zero-phonon line at $\sim637\,$nm) are used for optical excitation and readout. Continuous-wave 532\,nm green laser excitation pumps population from $^{3}\mathrm{A}_{2}$ to $^{3}\mathrm{E}$. Relaxation back to the ground state occurs either via radiative (optically allowed) decay or via spin-selective non-radiative decay through intermediate singlet states, which gives rise to spin-dependent fluorescence contrast. Dashed arrows indicate non-radiative or spin-non-conserving pathways. Spin sublevels are labeled by their magnetic quantum numbers ($m_{S}=0,\pm1$). MW control enables coherent spin manipulation between these sublevels, and an external magnetic field causes Zeeman splitting of the $m_{S}=\pm1$ sublevels. \textbf{(b) $\{100\}$-cut diamond lattice.} The N$V$ center direction (orange arrow) along $\langle 111 \rangle$ and the applied magnetic field $B_z$ (surface normal)\,\cite{welter2022scanning}. \textbf{(c) ODMR spectra of the N$V$ center}; the upper plot shows the resonance at zero magnetic field ($B = 0$). The lower plot shows the Zeeman splitting under a finite magnetic field ($B > 0$), where the resonance splits into two peaks separated by $2\gamma B_{NV}$, where $\gamma$ is the gyromagnetic ratio (28.024\,GHz/T).}
    \label{fig:NV_energy_level}
\end{figure*}

\section*{Acknowledgments}

This work is supported by the Carl Zeiss Stiftung (HYMMS Project No. P2022-03-044), the EU project HEU-RIA-MUQUABIS-101070546, and the DFG project FKZ: SFB 1552/1 465145163. K.M.C.F. acknowledges support from the University of Washington Molecular Engineering and Materials Center, funded by the National Science Foundation (NSF) Materials Research Science and Engineering Centers (MRSEC) program (Grant No. DMR-2308979). T.B., F.K., and M.K. acknowledge support from the Deutsche Forschungsgemeinschaft (DFG, German Research Foundation) through SPP 2137 (Project No. 403502522) and TRR 173 Spin+X (Projects A01, A12, and B02). The authors also acknowledge funding from TopDyn. This project received funding from the European Research Council (ERC) under the European Union’s Horizon 2020 research and innovation program (Grant No. 856538, project ``3D MAGiC'', and Grant No. 101070290, project ``NIMFEIA''), as well as from the Marie Sklodowska-Curie
grant agreement No. 101119608 (``TOPOCOM''). A.W. acknowledges support from EU, project HEU-RIA-MUQUABIS-101070546, by
the DFG, project FKZ: SFB 1552/1465145163 and by the German Federal Ministry of
Research, Technology and Space (BMFTR) within the Quantumtechnologien program
via the DIAQNOS project (project no. 13N16455). In addition it has been supported by the Ministerium für Wirtschaft, Arbeit und Tourismus (InvestBW),
project VitalQ, BW811662170, by the Helmholtz Association project Quantum Sensing for Fundamental Physics (QS4Physics) from the Innovation pool of the research
field Helmholtz Matter, QuBattery and QuBiopsy. A. M. and A. W. acknowledge support from the Deutsche Forschungsgemeinschaft (DFG, German Research Foundation) through TRR 173 Spin+X (Projects A09 and B14). Finally, the authors thank Muhib Omar and Jonas Raabe for their valuable technical support and assistance with programming throughout the course of this work.

\section*{APPENDIX A: N$V$ Wide-field magnetic imaging}

N$V^{-}$ centers in diamond possess optically addressable electron spin states that can be initialized, coherently manipulated, and read out at room temperature~\cite{pfaff2014unconditional,doherty2013nitrogen}. Ensembles of near-surface N$V^{-}$ centers produce real-space magnetic field maps over tens of micrometers with sub-micrometer spatial resolution~\cite{rondin2014magnetometry,levine2019principles}. This enables spatially resolved and minimally invasive measurements of complex magnetic textures, including skyrmions. Such features are difficult to access using bulk techniques such as SQUID, vibrating sample magnetometry, or X-ray scattering~\cite{hayami2021topological,casola2018probing,morgenbesser2021cation}.

Fig.\,\ref{fig:NV_energy_level}\,(a) shows the energy-level diagram of the N$V^{-}$ center. In its electronic ground state, the N$V^{-}$ center forms a spin-triplet ($S=1$) $^{3}A_{2}$ state, with Zeeman sublevels $m_{s}=0,\pm1$ separated by a zero-field splitting of $D \approx 2.87\,\mathrm{GHz}$~\cite{doherty2013nitrogen}. Optical excitation with green light (typically $\lambda = 532\,\mathrm{nm}$) polarizes the spin into the $m_{s}=0$ state. Spin readout is achieved via spin-dependent PL, since the $m_{s}=0$ state exhibits a higher PL rate than the $m_{s}=\pm 1$ states~\cite{doherty2013nitrogen}. 

Applying a static magnetic field along the N$V$ axis [Fig.\,\ref{fig:NV_energy_level}\,(b)] lifts the degeneracy of the $m_{s}=\pm1$ states through the Zeeman effect. ODMR is observed by sweeping a MW field across the resonance frequencies while monitoring the PL signal~\cite{rondin2014magnetometry}. At zero magnetic field ($B=0$), a small residual splitting in the ODMR spectrum [Fig.\,\ref{fig:NV_energy_level}\,(c)] arises from crystal strain or local electric fields that lift the degeneracy of the $m_{s}=\pm1$ sublevels even in the absence of an applied field~\cite{doherty2013nitrogen,tamarat2006stark}. High-quality diamond samples exhibit sufficiently narrow line-widths to resolve the hyperfine interaction with the $^{14}$N nuclear spin ($I=1$), producing a characteristic triplet splitting of approximately $2.16\,\mathrm{MHz}$~\cite{doherty2013nitrogen}. In the present measurements, however, this hyperfine structure is not resolved, and the analysis as well as the schematic description consider only the electronic spin transitions.

These features of the ODMR spectrum form the basis for magnetic field sensing, as shifts in the resonance frequencies directly encode the local magnetic field experienced by the N$V^{-}$ centers. By measuring these shifts across an ensemble of near-surface N$V^{-}$ centers, spatially resolved magnetic field information can be reconstructed. Wide-field ODMR imaging of near-surface N$V$ ensembles therefore enables two-dimensional magnetic field mapping over tens to hundreds of micrometers with sub-micrometer spatial resolution~\cite{rondin2014magnetometry,levine2019principles}.

\begin{figure*}
    \centering
    \includegraphics[width=1\textwidth]{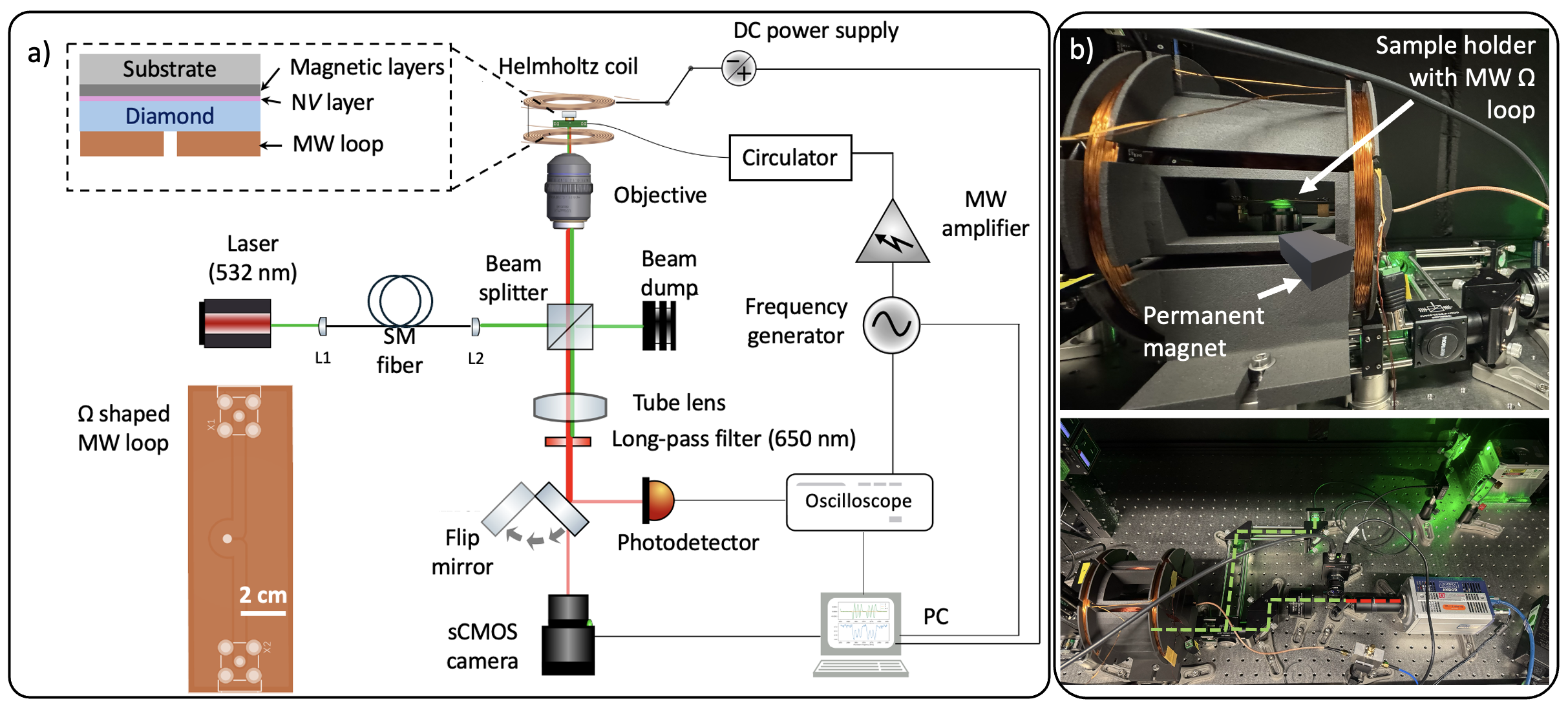}
    \caption{\textbf{Wide-field N$V$ magnetometry setup.} \textbf{(a)} Schematic of the optical and microwave setup. \textbf{(b)} Photographs of the experimental setup showing the sample holder with microwave loop and the optical table configuration.}
    \label{fig:setup}
\end{figure*}

\section*{APPENDIX B: Experimental setup}

The N$V$-based magnetic field measurements are performed using a custom-built inverted epifluorescence microscope [Fig.\,\ref{fig:setup}\,(a)]. A 532\,nm diode laser (Laser Quantum, gem 532\,nm) provides optical excitation of the N$V$ centers. To reduce pointing instability, the laser beam is coupled into a polarization-maintaining single-mode fiber (PMC-460Si-3-18E-150). A 50:50 beam-splitter cube (BSC) directs the excitation light toward the sample and separates the reflected and PL signals. Imaging is performed with an air objective (Newport M-20$\times$, NA = 0.40) with an effective focal length of 9\,mm.

The sample assembly consists of a custom flexible printed circuit board (PCB) of thickness 0.11\,mm incorporating an $\Omega$-shaped microwave stripline with a 3\,mm central aperture. The PCB is mounted on a three-dimensional translation stage for precise positioning. A diamond sensor containing a near-surface N$V$ layer is placed on the PCB, and the magnetic sample is positioned directly on top of the diamond to minimize the sensor–sample standoff distance. The diamond is oriented such that the N$V$ layer faces the magnetic film, maximizing magnetic sensitivity. The FOV is aligned with the center of the stripline loop to ensure uniform microwave excitation.

To filter the N$V$ photoluminescence, a 650\,nm long-pass filter (FEL0650, Thorlabs) is used. A flip mirror enables switching between imaging and point-detection configurations. Wide-field detection is performed using an sCMOS camera (Andor Zyla Wave 5.5\,MP) with a camera sensor pixel size of 6.5$\times$6.5\,$\mu$m$^{2}$, corresponding to an effective pixel size of approximately 325$\times$325\,nm$^{2}$ in the sample plane. For time-resolved or high-sensitivity measurements, an avalanche photodetector (APD440A, Thorlabs) is used, offering low dark counts and high quantum efficiency in the N$V$ emission band.

The performance of the wide-field N$V$ magnetometer is characterized by its magnetic field sensitivity, which is estimated from both shot-noise considerations and experimental noise measurements. The ideal shot-noise-limited magnetic field sensitivity of the magnetometer is calculated using the expression~\cite{barry2020sensitivity}

\begin{equation}
\eta_{\mathrm{CW}} = \frac{4}{3\sqrt{3}} \cdot \frac{h}{g_e \mu_B} \cdot \frac{\Delta \nu}{C_{\mathrm{CW}} \sqrt{R}},
\label{eq:sensitivity}
\end{equation}

\noindent
where $\Delta \nu$ is the ODMR linewidth, $R$ is the detected photon count-rate, and $C_{\mathrm{CW}}$ denotes the CW--ODMR contrast. 
The prefactor ${4}/{3\sqrt{3}}$ arises from the maximum slope of a Lorentzian resonance profile. 
Apart from the fundamental constants ($h$, $g_e$, and $\mu_B$), the sensitivity is governed by the ratio $\Delta \nu / C_{\mathrm{CW}}$, which reflects the sharpness of the ODMR resonance and its measurement contrast, as well as by $\sqrt{R}$. Using the experimentally determined parameters $\Delta \nu = 2.7\,\mathrm{MHz}$, $C_{\mathrm{CW}} = 4.5\%$, and $R = 1.85 \times 10^{8}\,\mathrm{counts/s}$, the ideal shot-noise-limited magnetic field sensitivity per pixel is estimated to be
\[
\eta_{\mathrm{CW}}^{\mathrm{ideal}} \approx 0.1\,\mu\mathrm{T}/\sqrt{\mathrm{Hz}}.
\]

Practically, however, the magnetic field sensitivity per pixel determined from the noise spectral density measurements yields
\[
\eta_{\mathrm{CW}}^{\mathrm{exp}} \approx 10.0-20.0\,\mu\mathrm{T}/\sqrt{\mathrm{Hz}},
\]
which is two orders of magnitude above the ideal shot-noise limit. To further quantify the origin of the discrepancy between the ideal and experimental sensitivity, we note that the dominant noise contributions arise from a combination of camera readout noise, laser intensity fluctuations, and microwave amplitude/frequency instability. Based on the measured noise spectral density, the effective magnetic field noise floor per pixel is on the order of $\sim 10\,\mu\mathrm{T}/\sqrt{\mathrm{Hz}}$. Among these contributions, technical noise sources (laser and microwave fluctuations) dominate over photon shot noise under the present operating conditions. This noise level sets the minimum detectable magnetic field variation in the reconstructed images. For an effective integration time (T) of $\sim 100$ ms per frequency point, the magnetic field noise per pixel is estimated as $\sigma_B \approx \eta/\sqrt{T} \approx 30\,\mu\mathrm{T}$. Given that the typical stray-field contrast associated with skyrmions is on the order of $0.3$--$1\,\mathrm{mT}$, this corresponds to a signal-to-noise ratio in the range $\mathrm{SNR} \sim 10$--$30$, which is sufficient to reliably resolve skyrmion features. Consequently, the extraction of skyrmion properties is not limited by sensitivity but rather by spatial resolution and segmentation accuracy.

However, the finite sensitivity contributes to uncertainty in the precise determination of skyrmion boundaries, particularly for low-contrast regions or near the edges of the skyrmions, where the magnetic field gradients are smaller. This leads to a small systematic uncertainty in the extracted geometrical parameters, which is mitigated in the present work by statistical averaging over large skyrmion ensembles.

\section*{APPENDIX C: Magnetic sensor}

For wide-field N$V$ magnetometry, we use a high-density N$V$ ensemble located near the diamond surface. A 100\,nm thick layer of $^{14}$N-doped, isotope-purified diamond (99.9\,\% $^{12}$C) is grown via chemical vapor deposition on an electronic-grade diamond substrate provided by Element Six. To create vacancies, the sample is implanted with 25\,keV He$^{+}$ ions at a dose of 10$^{12}$\,ions/cm$^{2}$, followed by vacuum annealing at 900$^{0}$C for 2\,h to form N$V$ centers. Subsequently, the sample undergoes annealing in O$_{2}$ at 425$^{0}$C for 2\,h to stabilize the charge state. The resulting N$V$ ensemble exhibits a density of approximately $1.2 \times 10^{17}\,\text{cm}^{-3}$, determined by comparing the fluorescence from this layer to that of near-surface single N$V$ centers in a reference sample. The diamond crystal has dimensions of $2 \times 2\,\text{mm}^2$ with a thickness of around 90–100\,$\mu$m~\cite{lenz2021imaging,kleinsasser2016high}.

\section*{APPENDIX D: External fields}

Magnetic domains are tuned using a bias magnetic field $B_z$ generated by a Helmholtz coil (radius = 4\,cm) [Fig.\,\ref{fig:setup}\,(b)], providing a field of approximately 1\,mT/A. The coil is driven by a computer-controlled DC power supply (Rigol DP-832), allowing magnetic field control with a resolution of approximately $10\,\mu\mathrm{T}$ during measurements. Microwave excitation for N$V$ magnetometry is provided by a signal generator (SRS SG384). The MW output is amplified using a 16\,W (+43\,dB) power amplifier (ZHL-16W-43+) and routed through a circulator (Pasternack CS-3.000) before delivery to the 50\,$\Omega$-terminated $\Omega$-shaped stripline integrated into the sample holder.

In addition to the OOP field, skyrmion nucleation requires an IP magnetic field component. This field is generated using a permanent magnet oriented parallel to the sample surface, producing an IP field of approximately 30\,mT at a distance of 1\,cm from the sample. The magnet is mounted on a lateral translation stage, allowing controlled and reversible application of the IP field. By adjusting its position, both the magnitude and direction of the IP field can be tuned without perturbing the optical alignment or overall experimental configuration.

\section*{APPENDIX E: Magnetic stand-off distance}

The finite N$V$–sample stand-off distance and the spatial extent of the N$V$ layer lead to a convolution of the measured magnetic field, effectively smoothing high-spatial-frequency features, slightly overestimating characteristic length scales, and reducing peak field amplitudes. To determine the N$V$–sample separation, we fit the measured stray-field profiles across DWs using the analytical two-edge stripe model~\cite{xu2025minimizing}, where a uniformly OOP magnetized domain is modeled as two anti-parallel bound currents at its edges and the resulting field is projected onto the N$V$ axis. The stand-off distance $z$ is obtained as a fitting parameter and represents an effective magnetic distance that includes both the physical gap and the finite N$V$ depth distribution, yielding a value of $z \approx 1$--$2\,\mu\mathrm{m}$, consistent with the average magnetic domain width of approximately $6\,\mu\mathrm{m}$. This extracted value is in good agreement with the experimentally observed magnetic field profiles.

\bibliography{FOCIS.bib}
\bibliographystyle{unsrt}

\end{document}